\documentclass[nohyper,11pt,letterpaper]{JHEP3}

\author{Valeri  Frolov\\
      Theoretical Physics Institute, Department of Physics\\
      University of Alberta, Edmonton, Canada, T6G 2J1\\
      E-mail: \email{frolov@phys.ualberta.ca}
}

\author{Dmitri Fursaev\\
      Bogoliubov Laboratory of Theoretical Physics,
      Joint Institute for Nuclear Research\\
      141980 Dubna, Moscow Region, Russia\\
      E-mail: \email{fursaev@thsun1.jinr.ru}
}

\author{Andrei  Zelnikov\\
      Theoretical Physics Institute, Department of Physics\\
      University of Alberta, Edmonton, Canada, T6G 2J1\\
      and \\
      Lebedev Physics Institute,
      Leninski pr. 53, Moscow  119 991, Russia\\
      E-mail: \email{zelnikov@phys.ualberta.ca}
}

\abstract{
We present a derivation of the entropy of black holes in induced
gravity models based on conformal properties of induced gravity
constituents near the horizon. The four-dimensional (4D) theory
is first reduced to a tower of two-dimensional (2D) gravities
such that each 2D theory is induced by fields with certain momentum
$p$ along the horizon. We demonstrate that in the vicinity of the
horizon constituents of the
2D induced gravities are described by conformal field theories (CFT)
with specific central charges depending on spin and non-minimal
couplings and with specific correlation lengths depending on the
masses of fields and on the momentum $p$. This enables one to use CFT
methods to count partial entropies $s(p)$ in each 2D sector.
The sum of partial entropies correctly reproduces the
Bekenstein-Hawking entropy of the 4D induced gravity theory. Our
results indicate that earlier attempts of the derivation of the
entropy of black holes based on a near-horizon CFT may have a
microscopic realization.
}


\preprint{{\tt hep-th/0302207}\\
Alberta-Thy-05-03
}

\title{CFT and Black Hole Entropy in \\Induced Gravity}

\begin{document}

\section{Introduction}
\setcounter{equation}0

The problem of a microscopic explanation of the Bekenstein-Hawking
entropy $S^{BH}$  of black holes \cite{Beken:72}--\cite{Hawking:75}
was a subject of intensive investigations over the last ten years.
This brought a number of very interesting results, especially in
string theory where $S^{BH}$ has been calculated by statistical
methods for certain types of black holes \cite{SV}  (for a review see
e.g. \cite{Peet}, \cite{DMW}). Despite of that progress the problem
is still far from its complete resolution and attracts a
lot of interest. The string theory has brought new stimulating
ideas, one of which was a role of two-dimensional conformal theory
(CFT) in the entropy counting.

The importance of the conformal symmetry was appreciated
after the computation made by Strominger
\cite{Strominger} who derived with its help the entropy of black holes in
2+1 dimensional asymptotically anti-de Sitter space-times \cite{BTZ}.
Later Carlip \cite{Carlip:99}--\cite{Carlip:02}
suggested a universal way
how to represent $S^{BH}$ in a statistical-mechanical
form by using a conformal symmetry of the gravitational action near
the black hole horizon  (see also work by
Solodukhin \cite{Solo:98}). The idea was that
a Poisson bracket realization of a near-horizon conformal symmetry
yields a Virasoro algebra with a central charge $c$ proportional to
$S^{BH}$. The entropy then can be derived
from the known expression for the degeneracy of states of CFT
with the given central charge at a fixed energy.

The approach \cite{Carlip:99}--\cite{Carlip:02} gives a statistical
{\it representation} of $S^{BH}$. It implies the existence of the
corresponding microstates but says nothing about their actual
physical meaning. Related to the fact that the microscopic theory is
not known, there is a certain ambiguity in definition of the
corresponding conformal theory. In particular, the derivation of the
entropy requires an introduction of an additional parameter  $l$, the
`size of the system'. Unfortunately, $l$, which can be considered as
a finite correlation length, cannot be fixed  from general
principles. For this reason it would be interesting to
understand how the ideas of \cite{Carlip:99}--\cite{Carlip:02} are
realized in simplified models of quantum gravity where microscopic
degrees of freedom are known.

There is a  class of such models known as the {\em induced
gravity theory}. The idea that gravity can be  induced by
quantum effects of matter fields goes back to Sakharov's work
\cite{Sakharov:68} (for a recent review of this subject see
\cite{Visser:02}). The microscopic degrees of freedom of induced
gravity are quantum fields responsible for generation of the Einstein
term in the low energy effective gravitational action. These fields are called
{\em constituents}.  The mechanism of the entropy generation in
the induced gravity is similar to the one proposed in
\cite{Sork,FrNo} where the Bekenstein-Hawking entropy was identified
with the entanglement entropy of quantum fields in the black hole
background. But there is  one important difference. Instead of the
entanglement of states of low energy physical fields, in the induced
gravity  the entropy is generated because of the entanglement  of
constituents \cite{Jacobson:94}.  This idea has been developed in
\cite{FFZ:96}, \cite{FF:97} where  a wide class of models of
induced gravity models was considered.

It should be emphasized that the induced gravity, at least in
its present form, is not a fundamental theory of quantum gravity. It
allows one to reproduce correctly tree-level results of the standard
Einstein gravity, but it does not give finite results for higher loop
quantities. Nevertheless in studying the black hole entropy problem the induced
gravity can serve as a convenient phenomenological model. Its main
characteristic properties  are the following: (i) the low-energy
gravity is an induced phenomenon, (ii) the underlying theory is free
of one-loop ultraviolet divergences, and (iii) there exist
microscopic degrees of freedom connected with the states of the
constituents which play the role of internal degrees of freedom.  One
can expect that any complete self-consistent theory of quantum
gravity must possess these properties. It is certainly true for the
string theory which is now usually considered as  candidate for this
role. There are indications that the microscopic explanation of the
black hole entropy in induced gravity may be  similar to the origin
of the entropy in open string theory \cite{MHS}. Thus, without
pretending to be as fundamental as the string theory, the induced
gravity may provide us with simple tools for
developing the intuition about the quantum gravity.

The constituents in models considered in \cite{FFZ:96}, \cite{FF:97}
are massive non-interacting scalar and
spinor quantum fields $\phi_k$. They propagate on a
classical background with metric $g_{\mu\nu}$ and
the action for $g_{\mu\nu}$ is defined as the
effective action $\Gamma[g]$ of the theory,
\begin{equation}\label{i.0a}
e^{i\Gamma[g]}=\int [D\phi]~e^{i\sum_k I[\phi_k,g]}~,
\end{equation}
where $I[\phi_k,g]$ is the action of the $k$-th field. The masses
of constituents are assumed to be of the order of the Planck mass
$m_{Pl}$.
In the low energy limit when the curvature $R$ of the background
is much smaller than $m^2_{Pl}$, the gravitational action coincides
with the Einstein-Hilbert action
\begin{equation}\label{i.0b}
\Gamma[g]\simeq {1 \over 16\pi G}\int \sqrt{-g} d^4x R[g]~.
\end{equation}
To make the models self-consistent the parameters of the constituents
are chosen
to eliminate the cosmological constant and cancel
ultraviolet divergences in the induced Newton constant $G$. To
satisfy the latter requirement one has to assume that some scalar
constituents are non-minimally coupled, i.e., include terms like $\xi
R\phi^2$ in the Lagrangian. In four-dimensional theories the
constraints leave a logarithmic divergence in a term in the effective
action which is quadratic in curvature. This, however, causes no
problem for studying entropy of  Schwarzschild and Kerr black holes
which are Ricci flat.

Note that the local temperature, as measured by an observer at
rest, infinitely grows near the black hole horizon.  Therefore, the
induced gravity constituents in the region close to the horizon
are effectively massless and conformally invariant. In the present
paper we investigate the role of conformal properties in derivation
of $S^{BH}$ in induced gravity. This enables one to test the idea of
\cite{Carlip:99}--\cite{Carlip:02} on concrete models where the
microscopic degrees of freedom are known. Our results suggest a
possible answer to some questions appearing in
\cite{Carlip:99}--\cite{Carlip:02}, like, for example, the origin of
a finite correlation length $l$. They also indicate that general
arguments when applied to a specific theory may need considerable
modification.

The paper is organized as follows. In section 2 we discuss
general properties of induced gravity models and the dimensional
reduction of these models. In particular, we show that after the
dimensional reduction  the induced gravity in the region near the
black hole horizon is reduced to a set two-dimensional (2D)
theories. These reduced theories describe  constituents with a
fixed momentum $p$ along the horizon. The four-dimensional (4D)
theory is recovered by summing over contributions of all
momenta. In this representation  the 4D induced constant
$G^{-1}$ is the sum of 2D gravitational couplings $G_2(p)$ at
different momenta $p$. This observation enables one to study 2D
induced gravities instead of 4D gravity.

Such theories are not conformally invariant. The conformal invariance
is broken by the masses and non-minimal couplings. We analyze the
role of these effects in section 3 and conclude that in the
near-horizon region each constituent can be related to a conformal
theory with some central charge and a finite correlation length. A
non-minimally coupled field with coupling $\xi$ results in a CFT with
a central charge $1-6\xi$.

In section 4, by taking into account that
at each momentum $p$ we have a set of CFT's with different charges
and different sizes we compute a "partial entropy" $s(p)$. The sum of
$s(p)$ over all momenta reproduces the correct value of $S^{BH}$. The
method of derivation of partial entropies is similar to computations
in \cite{Carlip:99}--\cite{Carlip:02}. We finish the paper by
summarizing our results and discussing remaining problems in
section 5. We also discuss a possible relation between the black hole
entropy and the information loss about the degrees of freedom under
the Wilson renormalization group flow on the space of 4D induced
gravity theories. Appendix A is devoted to properties of 2D induced
gravities obtained under the dimensional reduction. In Appendix B we
show that our derivation of the entropy works for black holes in
3-dimensional space-times.

\section{Induced gravity}
\setcounter{equation}0

\subsection{Models and general properties}

Induced gravity models may possess different types of the
constituent fields. In order to make the consideration more concrete
we use  the special model suggested in \cite{FFZ:96}. However, most
of the results of this paper can be easily generalized.

The model consists of $N_s$ scalar constituents $\phi_s$
with masses $m_s$, some
of the constituents being non-minimally coupled to the background
curvature with corresponding couplings $\xi_s$, and $N_d$ Dirac fields
$\psi_d$ with masses $m_d$. The corresponding actions
in (\ref{i.0a}) are
\begin{equation}\label{1.8}
I[\phi_s, g]=-\frac 12 \int d^4x \sqrt{-g} \left[(\nabla \phi_s)^2+
\xi_s R\phi_s^2+m_s^2\phi_s^2\right]~,
\end{equation}
\begin{equation}\label{1.9}
I[\psi_d, g]=\int d^4x \sqrt{-g}~ \bar{\psi}_d(i\gamma^\mu
\nabla_\mu+m_d)\psi_d~.
\end{equation}

We impose the following constraints on parameters
of the constituents:
\begin{equation}\label{1.1}
p(0)=p(1)=p(2)=p'(2)=0~,
\end{equation}
\begin{equation}\label{1.2}
q(0)=q(1)=0~,
\end{equation}
where
\begin{equation}\label{1.3}
p(z)=\sum_s m_s^{2z}-4\sum_d m_d^{2z}~,\hskip 1cm
q(z)=\sum_s m_s^{2z}(1-6\xi_s)+2\sum_d m_d^{2z}~.
\end{equation}
Constraints (\ref{1.1}) serve to eliminate
the induced cosmological constant while conditions (\ref{1.2})
enable one to get rid of the ultraviolet divergences
in the induced Newton constant $G$. It is  the second set of
conditions that will be important for our analysis of black hole
entropy.
Given (\ref{1.2}) $G$
is defined by formula
\begin{equation}\label{1.4}
{1 \over G}={1 \over 12\pi}q'(1)=
{1 \over 12\pi}\left[ \sum_s (1-6\xi_s)m_s^2\ln m_s^2 +2\sum_d
m_d^2\ln m_d^2\right]~.
\end{equation}
The Bekenstein-Hawking entropy of a black hole with the horizon
area $\cal A$
is given
by the usual formula
\begin{equation}\label{1.5}
S^{BH}={1 \over 4G} {\cal A}
\end{equation}
with $G$ defined in (\ref{1.4}).
Because $G$ is explicitly known one can prove that
\begin{equation}\label{i.1}
S^{BH}=S-Q~.
\end{equation}
Here $S$ is a statistical-mechanical entropy of the constituents
thermally distributed at the Hawking
temperature in the vicinity of the horizon .
The quantity
$Q$ is a quantum average of a Noether charge defined on
the bifurcation surface of the horizon by Wald's method \cite{Wald:93}.
The appearance of $Q$
is related to the fact that induced gravity models require
non-minimally coupled constituents.
Equation (\ref{i.1}) is universal: it is valid for different models
including those with vector constituents \cite{FF:98v}
as well as for different kinds of black holes,
rotating \cite{FF:99kerr} and charged
\cite{FF:99ch}, in different space-time
dimensions.

The physical reason of subtracting $Q$ in (\ref{i.1}), as was explained
in \cite{FF:97}, is related to two inequivalent definitions
of the energy in the black hole exterior. One definition, $H$,
is the canonical energy or the  Hamiltonian. The other definition,
$E$, is the energy expressed in terms of the stress-energy tensor
$T_{\mu\nu}$ which is obtained by variation of the action over the
metric tensor.  The two energies correspond to different
properties of a black hole. $H$ corresponds to evolution
of the system along the Killing time and for this reason
the operator $H$ in quantum theory is used for constructing the density
matrix which yields the entropy $S$ in (\ref{i.1}).
On the other hand, $E$ is related to thermodynamical properties of
a black hole. If the black hole mass measured at infinity is
fixed the change of the entropy $S^{BH}$ caused by the change of
the energy $E$ of fields in black hole exterior is
\begin{equation}\label{i.2}
\delta S^{BH}=-T_H \delta E~,
\end{equation}
where $T_H$ is the Hawking temperature of a black hole.
The reason why $E$ and $H$ are not equivalent is in the existence of
the horizon. The two quantities
being integrals of metrical and canonical stress tensors
differ by a total derivative. This difference
results in a surface term on the bifurcation surface of the
horizon. This surface term is not vanishing because the horizon is
not a real boundary and the only requirement for fields in this
region is regularity. One can show \cite{F:99n} that the boundary
term is the Noether charge $Q$ appearing in (\ref{i.1}). More
precisely,
\begin{equation}\label{i.3}
E=H-T_H Q~.
\end{equation}
According to (\ref{i.2}) the black hole entropy is related with  the
distribution over the energies $E$ of the induced gravity constituents.
Hence, the subtraction of $Q$ in (\ref{i.1}) accounts for the difference
between $E$ and $H$ in (\ref{i.3}).

It should be noted, however, that an explicit calculation of the
black hole degeneracy for a given mass $M$ which is connected with
the distribution of the constituent field states over the
energies $E$ is a problem. Two suggestions how it can be done are
discussed in \cite{FF:97} and \cite{F:99}. The difficulty is that in
quantum theory a non-zero value of $Q$ in (\ref{i.1}) is ensured by
the modes which, from the point of view of a Rindler observer, have
vanishing frequencies, the so-called soft modes.

\subsection{Dimensional reduction}

In what follows we assume that the black hole mass is much larger that
the Planckian mass and use a classical Schwarzschild metric to
describe it. Moreover,
since the black hole entropy is related to the behavior of the
constituents near the horizon, the curvature effects can be neglected
and we use the Rindler
approximation.
In this approximation the metric is
\begin{equation}\label{1.6}
ds^2=-\kappa^2 \rho^2 dt^2 +d\rho^2+dy_1^2+dy_2^2~,
\end{equation}
where $\kappa$ is the surface gravity and $y_j$ are coordinates
on the horizon.
In what follows we first discuss the induced gravity theory
on space-times with more general metrics
\begin{equation}\label{1.7a}
ds^2=dl^2+dy_1^2+dy_2^2~,
\end{equation}
\begin{equation}\label{1.7b}
dl^2=\gamma_{\alpha\beta}dx^\alpha dx^\beta~,
\end{equation}
where $\gamma_{\alpha\beta}$ in (\ref{1.7a}) is a metric tensor
of a 1+1 dimensional
Lorenzian space-time ${\cal M}_2$.
This generalization allows us to study conformal
properties of the dimensionally
reduced theories.

In this setting the dynamics of the constituents is essentially
two-dimensional which can be easily seen if we use the
Fourier decomposition in $y$-plane and define
\begin{equation}\label{1.14}
\phi_{s,{\bf p}}(x)={1 \over 2\pi a}\int d^2y~
e^{-i\bf{py}}\phi_s(x,y)~,\hskip 1cm
\psi_{d,{\bf p}}(x)={1 \over 2\pi a}\int d^2y~
e^{-i\bf{py}}\psi_d(x,y)~,
\end{equation}
where $\bf p$ is a momentum along the horizon,
${\bf p}{\bf y}=p_i y^i$.
To avoid volume divergences related to the infinite size of the
horizon we  assume that the range of coordinates $y^i$
is restricted,
$-a/2\leq y^i \leq a/2$. This means that the horizon area
${\cal A}$ is finite and equal to $a^2$.

With these definitions actions (\ref{1.8}), (\ref{1.9})
on space-time (\ref{1.7a}) can be written
in the form
\begin{eqnarray}\label{1.10}
I[\phi_s, g]&=&\sum_{\bf p}I[\phi_{s,{\bf p}}, \gamma]~,\hskip 2cm
I[\psi_d, g]=\sum_{\bf p}I[\psi_{d,{\bf p}}, \gamma]~,\\
\label{1.11}
I[\phi_{s,{\bf p}}, \gamma]&=&-\frac 12 \int d^2x \sqrt{-\gamma}
\left[(\nabla \phi_{s,{\bf p}})^2+
\xi_s {\cal R}~\phi_{s,{\bf p}}^2+m_s^2({\bf p})\phi_{s,{\bf
p}}^2\right]~,\\ \label{1.12}
I[\psi_{d,{\bf p}}, \gamma]&=&\int d^2x \sqrt{-\gamma}~
\bar{\psi}_{d,{\bf p}}(i\gamma^\alpha
\nabla_\alpha+m_d+\gamma^j p_j)\psi_{d,{\bf p}}~,
\end{eqnarray}
where ${\cal R}$ and $\nabla$'s are, respectively, the scalar curvature
and covariant derivatives
on ${\cal M}_2$ and
\begin{equation}\label{1.15}
m_s^2({\bf p})=m_s^2+{\bf p}^2~.
\end{equation}
The fact that for a spinor field $\psi_{d,{\bf p}}$ the mass term
is described by a matrix
$m_d+\gamma^j p_j$ does not create problem since
the propagator in this theory has a pole at
$m_d^2({\bf p})=m_d^2+{\bf p}^2$.
The quantity $m^2_k({\bf p})$ (for $k=s$ or $d$) can be interpreted
as the energy square of a particle
with mass $m_k$ and the transverse momentum ${\bf p}$.
Thus, as follows from (\ref{1.10}), (\ref{1.11}) each induced gravity
constituent in the process of dimensional reduction produces
a tower of 2D
fields with masses equal to transverse energies of 4D fields.

Let us now discuss how  to derive the induced Newton constant
(\ref{1.4}) in terms of parameters of two dimensional fields.
In induced gravity ultraviolet divergencies are not just
truncated, but cancel themselves because of constraints.
Therefore, in induced gravity the 4D effective action $\Gamma[g]$
is the sum of actions of dimensionally reduced models
(see, e.g., \cite{FrSuZel:00}). So, on space-time (\ref{1.7a})
\begin{equation}\label{1.16}
\Gamma[g]=\sum_{\bf p} \Gamma_2[\gamma,|{\bf p}|]
={a^2 \over 4\pi}\int^\infty_\sigma \Gamma_2[\gamma,p] dp^2 ~.
\end{equation}
Here $\Gamma_2[\gamma,p]$ is the effective action of
two-dimensional gravity induced by constituents with the momentum
$p=|{\bf p}|$.
In the last equality in (\ref{1.16})
we assumed that parameter $a$ is large and replaced
the sum over $\bf p$  by the integral over $p$.
The coefficient $a^2 /(4\pi)$ is related to
the number of modes with momentum $p^2$ in an interval
$\Delta p^2$. The lower integration limit $\sigma$ in (\ref{1.16})
is proportional to $a^{-2}$ and is determined by the smallest
momentum of a particle in the region of the horizon with area
${\cal A}=a^2$. At infinite $a$, $\sigma$ vanishes.

The two-dimensional action can be easily calculated in the
limit when the curvature ${\cal R}[\gamma]$ of 2D space-time with
metric (\ref{1.7b}) is smaller than $m_k^2(\bf p)$. One finds
\begin{equation}\label{1.17}
\Gamma_2[\gamma,p]\simeq{1 \over 4G_2(p)} \int\sqrt{-\gamma} d^2x
~\left({\cal R}+2\lambda_2(p)\right)~.
\end{equation}
Here $G_2(p)$ and $\lambda_2(p)$ are, respectively,
the gravitational coupling
and cosmological constant of the 2D theory,
\begin{eqnarray}\label{1.18}
{1 \over G_2(p)}&=&-{1 \over 12\pi}
\left[ \sum_s (1-6\xi_s)\ln (m_s^2+p^2) +2\sum_d
\ln (m_d^2+p^2)\right]~,\\
\label{1.19}
{\lambda_2(p) \over G_2(p)}&=&{1 \over 4\pi}
\left[ \sum_s (m_s^2+p^2)\ln (m_s^2+p^2) -4 \sum_d (m_d^2+p^2)
\ln (m_d^2+p^2)\right]~.
\end{eqnarray}
Note that the same constraints which provide
finiteness of the four-dimensional couplings also
eliminate the divergences of $G_2(p)$ and $\lambda_2(p)$.

The four-dimensional Newton constant $G$ can be found by
summation over momenta in (\ref{1.16}). One gets
\begin{equation}\label{1.20}
\Gamma[g]={1 \over 16\pi G}\int \sqrt{-g}d^4x R[g]~,
\end{equation}
where $R[g]={\cal R}[\gamma]$ because
the 4D action is considered on the space-time
(\ref{1.7a}), (\ref{1.7b}).
We also put here $a^2=\int dy_1 dy_2$.
The constant $G$ is defined as
\begin{equation}\label{1.21}
{1 \over G}=\lim_{p \rightarrow 0}{1 \over G(p)}~,\hskip 2cm
{1 \over G(p)}=\int_{p^2}^\infty {d{\tilde p}^2 \over G_2({\tilde p})}~,
\end{equation}
and coincides with (\ref{1.4}).
Again, the integral in (\ref{1.21}) is finite due to constraints
(\ref{1.2}).

\subsection{Running gravitational couplings}

Equation (\ref{1.21}) can be considered as a formal
representation of the induced Newton constant.
In what follows, however, we would like to move further and
treat the two dimensional field models at
any momentum $p$ as real physical
theories\footnote{It should be noted that
the idea of two-dimensional reduction of quantum gravity at
high energies is not new. It also appears under discussion of scattering of
particles with energies at the center of mass comparable or larger than
the Planckian energy, see \cite{Hooft}, \cite{Verlinde}.}.
This, certainly,
imposes further restrictions on parameters of the constituents
dictated by physical requirements.
Because (\ref{1.21}) includes particles with arbitrary large
momenta $p$ we are forced to make some guesses about the
theory beyond the Planckian energies.
(Note that the contribution of momenta much
larger than masses of all constituents asymptotically vanish because
of constraints.)
It is natural to assume that at any $p$
the 2D induced gravities have strictly
positive gravitational couplings $G_2(p)$.
As one of the arguments note that only in this case
the Bekenstein-Hawking entropy of a back hole in such 2D
theories is positive.
This assumption is closely related to properties of the
other parameter, $G(p)$, in (\ref{1.21}). $G(p)$
has a meaning of a running Newton coupling constant
in the
theory with the ultraviolet cutoff at the transverse momentum
$p$.
To demonstrate this,
let us suppose that
all constituents have masses of the order of $m_{Pl}$ and that the
background curvature $|{\cal R}|\simeq m^2_{Pl}+p^2$.
Then the low-energy approximation (\ref{1.17}) for induced actions
is valid only for constituents with momenta larger than chosen value $p$.
In this case the gravitational coupling
of the four-dimensional theory is $G(p)$ because it is
determined only by contributions of fields with momenta above $p$.

Interpretation of $G(p)$ as a running coupling also follows from the
definition of the effective action in the Wilson renormalization group
approach, see \cite{WK} (and chapter 12 of \cite{PS} for a good
introduction). In standard Wilson approach the effective action at the scale
$p$ is defined by integrating in the functional integral over all modes with
momenta larger than $p$  (one uses the Euclidean formulation). The
distinction of our case is that the momentum is considered as the transverse
one. After such integration one gets the induced gravity theory with
constant $G(p)$ plus quantum theory of "low-energy" constituents with
transverse momenta smaller than $p$.

If $G(p)$ is a running coupling, $G_2(p)$ is
related to the beta-function of the theory because
\begin{equation}\label{1.22}
{1 \over G_2(p)}=-{\partial \over \partial p^2}{1 \over G(p)}~.
\end{equation}
When $G_2(p)$ is strictly positive, the
four-dimensional coupling $G(p)$ monotonically increases
and the gravity gets stronger and stronger at high energies.
It is this  sort of the behavior which one may expect
in quantum gravity and it is what one finds in the Wilson renormalization
group.
We return to discussion of this point in the last section.

In Appendix A we show that there is a class of induced
gravity models with a softly broken supersymmetry where the
cosmological constant $\lambda_2(p)$ is zero,  while the
gravitational coupling $G_2(p)$ is positive for all $p>0$.

\section{Conformal properties near the horizon}
\setcounter{equation}0

\subsection{Effects related to masses}

Let us discuss properties of two-dimensional fields
$\phi_{s,{\bf p}}$, $\psi_{d,{\bf p}}$. If the fields were massless,
and, additionally, $\phi_{s,{\bf p}}$ were minimally coupled,
the classical actions (\ref{1.11}), (\ref{1.12}) would be conformally
invariant. In quantum theory the conformal invariance would be broken
by anomalies.

It is convenient to proceed further in the Euclidean version of the theory
by assuming that Euclidean analog of ${\cal M}_2$ has a topology of
a disk and the Euclidean time $\tau$ (related to the Rindler time)
is a periodic coordinate.
The Euclidean actions have the same form as Lorenzian functionals
(\ref{1.11}), (\ref{1.12}) except that the scalar action has a
different sign and factor $i$ by the Dirac operator in (\ref{1.12})
should be omitted.  With definition of the stress-energy tensor
$$
T_{\alpha\beta}=-{2 \over \sqrt{\gamma}}{\delta I \over \delta
\gamma^{\alpha\beta}}
$$
one has
\begin{equation}\label{2.1}
T_{\alpha\beta}=-\left(\nabla_\alpha\phi\nabla_\beta\phi-
\frac 12 ((\nabla \phi)^2+m^2+\xi
\tilde{R}~\phi^2)\gamma_{\alpha\beta}+\xi (\gamma_{\alpha\beta}
\nabla^2-\nabla_\alpha\nabla_\beta )\phi^2\right)~,
\end{equation}
for a 2D scalar constituent with a mass $m$
(which depends on $p$) and a non-minimal coupling $\xi$.
For a 2D spinor constituent (with or without mass)
\begin{equation}\label{2.2}
T_{\alpha\beta}={1 \over 4}
\left(\nabla_\alpha\tilde{\psi}\gamma_\beta\psi+
\nabla_\beta\tilde{\psi}\gamma_\alpha\psi
-\tilde{\psi}\gamma_\alpha \nabla_\beta\psi-
\tilde{\psi}\gamma_\beta \nabla_\alpha\psi
\right)~.
\end{equation}
Here $\tilde{\psi}$ is an analog of conjugated spinor
which in the Euclidean
theory is a variable independent of $\psi$.

Consider our theory on the Rindler space and make the Wick rotation.
We obtain the Euclidean theory on 2D space
\begin{equation}\label{2.3}
dl^2=\kappa^2\rho^2d\tau^2+d\rho^2=dz d\bar{z}~,
\end{equation}
$0\leq \tau \leq 2\pi/\kappa$ and $0\leq \rho \leq \rho_+$
where $\rho_+$ is some radius. The complex coordinates
are introduced as usually, $z=x_1+ix_2$, where
$x_1=\rho\cos \kappa \tau$, $x_2=\rho\sin \kappa \tau$.
If the constituents are massless and minimally coupled
the scalar action (\ref{1.11}) takes form
\begin{equation}\label{2.5}
I[\phi]=2\int d^2z~ \partial \phi \bar{\partial}\phi
\end{equation}
where {\bf $dz^2=dx_1dx_2$,} $\partial=\partial/\partial z= \frac 12(\partial_1-i\partial_2)$,
$\bar{\partial}=\partial/\partial \bar{z}$.
The integral (\ref{2.5}) is invariant under
the conformal transformation
\begin{equation}\label{2.4}
z=f(w), ~~\bar{z}=\bar{f}(\bar{w})~,
\end{equation}
\begin{equation}\label{2.4a}
\phi(z,\bar{z})=
\phi'(w,\bar{w})~.
\end{equation}
To write transformations in the spinor theory it is convenient
to define
the components of spinors in (\ref{1.12}) as follows
$\psi^{T}=(\bar{b}_1,a_1,\bar{b}_2,a_2)$,
$\tilde{\psi}=(b_1,\bar{a}_1,b_2,\bar{a}_2)$, where
$\bar{b}_j,\bar{a}_j$ and $b_j, a_j$ are independent variables.
If we neglect
the masses and transverse momenta the action (\ref{1.12})
for a particular constituent has the form\footnote{We use
representation
where $\gamma^\alpha=(\sigma^1 \times \sigma^\alpha)$, $\alpha=1,2$
and $\sigma$'s are the Pauli matrices.}
\begin{equation}\label{2.6}
I[\psi]=\int d^2x ~\tilde{\psi}\nabla^\alpha \gamma_\alpha\psi
=\sum_{j=1,2}\int d^2z \left[b_j \bar{\partial} a_j+
\bar{b}_j \partial \bar{a}_j\right]~.
\end{equation}
It is invariant under conformal transformations
of coordinates (\ref{2.4}) when the spinor components transform as
\begin{equation}\label{2.7a}
b_j(z,\bar{z})=(\partial f)^{-1/2}
b'_j(w,\bar{w})~,\hskip 1cm a_j(z,\bar{z})=(\partial f)^{-1/2}
a'_j(w,\bar{w})~,
\end{equation}
\begin{equation}\label{2.7b}
\bar{b}_j(z,\bar{z})=(\bar{\partial}\bar{f})^{-1/2}
\bar{b}_j'(w,\bar{w})~,\hskip 1cm \bar{a}_j(z,\bar{z})=(\bar{\partial}\bar{f})^{-1/2}
\bar{a}_j'(w,\bar{w})~,
\end{equation}
where $\partial f=\partial_w f$.
Let us consider now the holomorphic components of the
stress-energy
tensor\footnote{The commonly used definition of $T(z)$ differs from
our definition by the factor $2\pi$.},
$T(z)=T_{zz}(z)$, for scalars and spinors, respectively,
\begin{equation}\label{2.8}
T(z)=-(\partial \phi)^2 ~,
\end{equation}
\begin{equation}\label{2.9}
T_j(z)=\frac 14 (\partial b_j a_j-b_j \partial a_j)~.
\end{equation}
(The anti-holomorphic components are defined analogously).
It is well-known that in quantum theory the renormalized
components transform as (see, e.g., \cite{Cardy})
\begin{equation}\label{2.8a}
T'(w)=(\partial f)^2 T(z)+{1 \over 2\pi} A_f(w)~,
\end{equation}
\begin{equation}\label{2.9a}
T'_j(w)=(\partial f)^2 T_j(z)+{1 \over 2\pi} A_f(w)~,
\end{equation}
\begin{equation}\label{2.10}
A_f(w)={2\partial^3f \partial f -3 (\partial ^2f)^2
\over 24(\partial f)^2}~.
\end{equation}
An anomalous term $A_f(w)$ appears in the transformation law
because renormalization procedure requires subtracting
divergent terms which are
not conformally invariant. As follows from (\ref{2.8a}), (\ref{2.9a}),
the conformal algebra in quantum theory has a central extension.
The corresponding central charge $c$ for each scalar field and
each $j$-th component of the spinor field equals unity (thus the total
central charge corresponding to a dimensionally reduced Dirac spinor
is $c=2$).

Let us now discuss what happens when fields are massive.
Consider first a scalar field with a mass $m$. Consider the correlation
function which determines the theory. In two dimensions
\begin{equation}\label{2.11}
\langle \phi(z,\bar{z}) \phi(z',\bar{z}') \rangle=
{1 \over 4\pi}\int_0^\infty {ds \over s}e^{-{d^2 \over 4s}-m^2s}=
{1 \over 2\pi}K_0(d^2m^2)~,
\end{equation}
where $d^2=|z-z'|^2$ and $K_\nu(x)$ is the Bessel function.
At small $x$ the Bessel function has the following asymptotics
\begin{equation}\label{2.12a}
K_0(x)=-\ln {x \over 2} +O(x^2 \ln x)~,
\end{equation}
while at large $x$
\begin{equation}\label{2.12b}
K_0(x)=\sqrt{\pi \over 2x}e^{-x}(1 +O(x^{-1}))~.
\end{equation}
Therefore, if the points $(z,\bar{z})$ and $(z',\bar{z}')$
are entirely inside the circle of the radius $R<m^{-1}$
the correlator is determined by the leading term in
(\ref{2.12a}),
\begin{equation}\label{2.13}
\langle \phi(z,\bar{z}) \phi(z',\bar{z}') \rangle=
-{1 \over 4\pi}\ln (m^2 |z-z'|^2)~,
\end{equation}
and it coincides up to an irrelevant constant with the correlator of
the massless field. On the other hand, if one of the points is
inside of the circle and the other is far outside, or the two points
are outside and far apart from each other the correlator
is exponentially small, see (\ref{2.12b}).

This analysis shows that if we are interested in the behavior
of the  system inside the circle with radius $m^{-1}$ we can consider
it as a massless theory. In general, the mass breaks conformal
invariance explicitly, but inside this circle the effect of the mass is
small and theory is conformal. The conformal properties
improve when one shrinks the circle toward its center.
The only feature related to the fact that our theory is obtained
as a result of the dimensional reduction is that mass is a function
of the transverse momentum $p$.
Thus, the radius is $R\sim (m^2+p^2)^{-1/2}$
and the larger the momentum $p$ the smaller
the region where theory is conformal.

Let us now discuss the case of spinor fields. If we keep mass and
momentum terms the reduced action for a spinor with a momentum $p$
has the form\footnote{We put $\gamma_{i=1}=(-i\sigma_2\times I_2)$
(where $I_2$ is the unit matrix) and
$\gamma_{i=2}=(\sigma_1\times \sigma_3)$.}
\begin{eqnarray}
I[\psi]&=&\int d^2x ~\tilde{\psi}
(\nabla^\alpha \gamma_\alpha\psi-ip_j\gamma^j+im)\psi
\nonumber \\
&=& \label{2.14}
\int d^2z \left[\sum_{j=1,2}\left(b_j \bar{\partial} a_j+
\bar{b}_j \partial \bar{a}_j+{im\over 2}
(\bar{a}_ja_j+\bar{b}_jb_j)\right)  \right.     \\
&& \left.\phantom{\sum_j}\hskip 40pt
+{\bar{\mu} \over 2}
(b_1\bar{b}_2-\bar{a}_2a_1)+
{\mu \over 2}
(\bar{a}_1a_2-b_2\bar{b}_1)
\right]~, \nonumber
\end{eqnarray}
where $\mu=\mu({\bf p})=p_1+ip_2$,
$\bar{\mu}=p_1-ip_2$. In conformal theory (\ref{2.6})
the only nonzero correlators are $\langle b_j a_j\rangle$
and $\langle \bar{b}_j \bar{a}_j\rangle$.
In theory (\ref{2.14}) new correlators appear. All of them are
easy to calculate. For instance, for the field
$b_1$ there are three non-trivial correlators:
\begin{equation}\label{2.15}
\langle b_1(z,\bar{z}) a_1(z',\bar{z}') \rangle=
-{1 \over \pi} \partial ~K_0(d^2 m^2({\bf p}))~,
\end{equation}
\begin{equation}\label{2.16}
\langle b_1(z,\bar{z}) \bar{b}_1(z',\bar{z}') \rangle=
-{im \over 2\pi}~ K_0(d^2 m^2({\bf p}))~,
\end{equation}
\begin{equation}\label{2.17}
\langle b_1(z,\bar{z}) \bar{b}_2(z',\bar{z}') \rangle=
{\mu({\bf p}) \over 2\pi}~ K_0(d^2m^2({\bf p}))~.
\end{equation}
where $d^2=|z-z'|^2$.
Suppose now that
both points are inside the circle with radius
$R\sim |m({\bf p})|^{-1}$. Then the correlator
can be approximated by that of the conformal theory,
see (\ref{2.13}). In this case $\langle b_1 a_1 \rangle$ is the
correlator of conformal fermions. Other correlators (\ref{2.16}),
(\ref{2.17}) are not zero in this limit. However, they are smaller
compared to $\langle b_1 a_1 \rangle$. Indeed, one can see that
$\langle b_1 a_1 \rangle \sim (z-z')^{-1}$, while
$\langle b_1 \bar{b}_1 \rangle \sim  m\ln|z-z'|$,
$\langle b_1 \bar{b}_2 \rangle \sim \mu \ln|z-z'|$. Thus, the new
correlators can be neglected once the points are inside the circle
and in this region fermions are well described by a conformal theory.

\subsection{Rindler entropy in 2D theory}

Let us discuss the entropy $s$ of a massless scalar field in
a two-dimensional Rindler space-time (see (\ref{1.6}))
\begin{equation}\label{3.1}
dl^2=-\kappa^2 \rho^2 dt^2+d\rho^2~.
\end{equation}
We define $s$  as the entropy
of Rindler quanta in some range of coordinate $\rho$,
$\epsilon \leq \rho \leq R$. For quanta at the temperature
$T_H={\kappa \over 2\pi}$ the entropy is known to be
\begin{equation}\label{3.2}
s={1 \over 6} \ln {R \over \epsilon}~.
\end{equation}
If we have $c$ fields
\begin{equation}\label{3.3}
s(c,R/\epsilon)={c \over 6} \ln {R \over \epsilon}~.
\end{equation}
Formula (\ref{3.3}) also holds for an arbitrary  CFT if constant $c$
in it is identified with the corresponding central charge.
Such a result can be  obtained
by using conformal properties of the free energy and
pointing out that $c$ is related to the conformal anomaly
\cite{Affleck}.

The entropy of Rindler particles can be also interpreted as the
entropy of entanglement between quantum
states inside and outside the horizon.
It is important to note that this entropy can be derived
using the well-known Cardy formula,
in a way which is parallel to the derivation of black hole entropy
in Carlip's approach
\cite{Carlip:99}--\cite{Carlip:02}. If the central charge $c$ is fixed,
$s(c,R/\epsilon)$ determines the degeneracy of
operators $L_0$, $\bar{L}_0$ (which generate translations along coordinates
$u=t-x$, $v=t+x$) in the state which corresponds to the Minkowski
vacuum.  We do not dwell here on how the Cardy formula can be used
to obtain (\ref{3.3}), the details can be found, e.g., in
\cite{FFGK:99}.

\bigskip

It follows from discussion of the preceding section that
result (\ref{3.3}) should not change if the field has a
non-zero mass $m$
provided that $R$ is smaller than the corresponding correlation length,
$R< m^{-1}$. This is easy to see by examining the equation
for a Rindler mode
$\phi_\omega(t,\rho)=e^{-i\omega t}\phi_\omega(\rho)$ with
frequency $\omega$,
\begin{equation}\label{2.20a}
((\kappa^{-1}\omega)^2-H^2)\phi_\omega=0~,
\end{equation}
\begin{equation}\label{2.20b}
H^2=-(\rho \partial_\rho)^2+\rho^2m^2~.
\end{equation}
The operator $H$ is a single-particle Hamiltonian.
According to (\ref{2.20a}), the closer
particle to the Rindler horizon $\rho=0$
the smaller the effect of its mass. The mass becomes important
when $\rho \sim m^{-1}$.

By taking into account that the normalized single-particle mode
is expressed in terms of the Bessel function,
\begin{equation}\label{2.22a}
\phi_\omega(\rho)=
{\sqrt{\sinh (\pi \omega/\kappa)} \over \kappa \pi^2}
K_{i\omega/\kappa}(m\rho)~,
\end{equation}
the free energy of the field in the region $\epsilon <\rho < R$
can be written in the form
\begin{equation}\label{2.22}
F(\beta,\epsilon,R)=\beta^{-1} \int_0^\infty \Phi(\omega,\epsilon,R)
~\ln\left(1-e^{-\beta\omega}\right)d\omega~,
\end{equation}
\begin{equation}\label{2.23}
\Phi(\omega,\epsilon, R)=2\omega \int_\epsilon^R
{d \rho \over \rho} |\phi_\omega(\rho)|^2
~.
\end{equation}
The quantity $\Phi(\omega,\epsilon,R)d\omega$ is the number
of energy levels in the interval $(\omega,\omega+d\omega)$.
If $R < m^{-1}$ and $\epsilon \rightarrow 0$
the spectral density (\ref{2.23}) coincides with the
spectral density of the massless theory in the same region
$\epsilon < \rho < R$.
However, for $R>m^{-1}$ the contribution
from the domain
$\rho>m^{-1}$ to the
integral in (\ref{2.23}) is exponentially suppressed
due to the behavior of the Bessel function.
This means that for $R>m^{-1}$ in the leading asymptotic
the free energy coincides with the free energy of massless theory
in the region $\epsilon < \rho < m^{-1}$.
Such a conclusion is supported by the direct calculation
which shows that for
$\epsilon \ll m^{-1} \ll R$
\begin{equation}\label{2.25}
\Phi(\omega,\epsilon,R)=-{1 \over \pi \kappa} \ln {m \epsilon \over 2}
+f(\omega,\epsilon)+\bar{f}(\omega,\epsilon)~,
\end{equation}
\begin{equation}\label{2.25a}
f(\omega,\epsilon)={i \over 4\pi \omega} { \Gamma(1+i\omega/\kappa)
\over \Gamma(1-i\omega/\kappa)}
\left({m \epsilon \over 2}\right)^{-2i\omega/\kappa}
-{1 \over 2\pi\kappa}\psi(1-i\omega/\kappa)~,
\end{equation}
where $\bar{f}$ denotes complex conjugated function and $\psi(x)$ is
the logarithmic derivative of the $\Gamma$ function $\Gamma(x)$.
In the limit $\epsilon \rightarrow 0$ the first term in the r.h.s.
in (\ref{2.25}) dominates over the last two ones. Thus, leaving in
the spectral density only the logarithmic term one gets from
(\ref{2.22})
\begin{equation}\label{2.26}
F(\beta,\epsilon,R)=-{\pi \over 6\beta^2 \kappa}\ln {1 \over m \epsilon}
~.
\end{equation}
The corresponding entropy computed at
the Hawking temperature $\beta^{-1}=\kappa/2\pi$ is
\begin{equation}\label{2.24}
s(R/\epsilon)={1 \over 6} \ln {1 \over m \epsilon}~,
\end{equation}
as was expected.

\subsection{Effects related to non-minimal couplings}

So far we discussed the effects of massive terms.
In addition to them there are other terms which break conformal
invariance. In our theory these are non-minimal couplings
in the scalar action (\ref{1.11}). These terms modify the
scalar field stress-energy tensor and its transformation
properties. Instead of (\ref{2.8}), (\ref{2.8a}) in case
of non-minimal couplings one has (see, e.g., \cite{LW:95})
\begin{equation}\label{2.18}
T(z)=-(\partial \phi)^2+2\xi((\partial \phi)^2+\phi \partial^2 \phi) ~,
\end{equation}
\begin{equation}\label{2.19}
T'(w)=(\partial f)^2 T(z)+(1-6\xi){1 \over 2\pi} A_f(w)
-\xi{1 \over 8\pi} (\partial \ln \partial f)^2~,
\end{equation}
where we neglected by mass terms.
Without the last term
in r.h.s. of (\ref{2.19}) this transformation law corresponds to a
conformal theory with an effective central charge $c=1-6\xi$.

Let us discuss conditions when the last term can be ignored.
It is easy to see that this can be done if the conformal
transformations are infinitesimal,
\begin{equation}\label{2.30}
z=f(w)=w+\varepsilon (w)~,\hskip 1cm |\varepsilon (w)|\ll 1~.
\end{equation}
In this case variation of the stress-energy tensor
under transformation (\ref{2.19}) in the leading order in
$\varepsilon$ is
\begin{equation}\label{2.31}
\delta T(w)=T'(w)-T(w)= \varepsilon (w) \partial T(w)+
2\partial\varepsilon(w) T(w) +{1-6\xi \over 24\pi}
\partial^3\varepsilon(w) +O(\varepsilon^2)
~.
\end{equation}
The last term in (\ref{2.19}) gives contribution proportional
to $(\partial\varepsilon)^2$ which is not important
at  this order.

Note that it is the linear order of the variations
of the conformal stress-energy tensor which is
needed to infer the structure of the Virasoro algebra.
Therefore, when infinitesimal conformal
transformations are considered, a 2D scalar field $\phi$
with a non-minimal coupling $\xi$
is equivalent to a conformal theory with the effective
central charge $c=1-6\xi$. The concrete realization of this
effective conformal theory in terms of the initial field $\phi$
may be non-trivial\footnote{It should be noted that the
generator of the conformal transformation $\delta T$ in
(\ref{2.31}) is not the operator $T$ itself but only its part
which does not depend on the parameter $\xi$.}
but it is not important for the computation of the entropy.
What really matters is the value of the central charge which
enables one to fix uniquely the form of the free energy
by using conformal properties of the effective action
\cite{Affleck} and use formula (\ref{3.3}) for the entropy.

The relevance of infinitesimal conformal transformations
for computation of the entropy can be also seen
from the following reasonings \cite{LW:95}.
In the Euclidean theory the entropy
can be derived in a geometrical way by using the conical singularity
method. To this aim one has to consider
the transformation $f(w)=w^\alpha$ with
$\alpha ={2\pi \over \beta}$. Then in $(w,\bar{w})$--plane
there is a conical
singularity with the deficit angle $2\pi-\beta$. The parameter $\beta$
corresponds to an inverse temperature and the limit
$\beta=2\pi$ corresponds to the Hawking temperature, or to a Minkowski
vacuum in case of the Rindler space-time. To derive the
entropy one has to take the Euclidean action off-shell,
find its derivative over $\beta$ and after that
go to limit $\beta=2\pi$. It is easy to see that at small angle
deficits $A_f(w)\sim (2\pi-\beta)$ and the corresponding term
in (\ref{2.19})
is related to contribution to the entropy proportional to
the effective central charge $c=1-6\xi$. As for the last
term, its contribution
vanishes because $(\partial \ln \partial f)^2\sim (2\pi-\beta)^2$.

\subsection{Negative central charge contributions}

The constraints of induced gravity models require that at least
some of effective central charges $c=1-6\xi$ were negative.
For $c<0$ formula (\ref{3.3}) results in a negative entropy.
Typically CFT's with negative central charges correspond to ghosts.
The ghosts appear in gauge theories when the
Hilbert space is enlarged during the
quantization. The ghosts  do have negative entropy because it
should
compensate contribution of the extra degrees of freedom in the
enlarged Hilbert space. However, if the system  is unitary its
total entropy is always positive.

There is some similarity between induced gravity models and gauge
theories in the following sense \cite{FF:97}. As we discussed in
section 2.1, the non-minimal couplings result in modification of the
energy of the system $E$ (adding a surface term $T_H Q$ at the
horizon) but do not change the canonical energy $H$, see (\ref{i.3}).
In quantum theory the value of $Q$ is determined by the soft modes,
i.e., modes with vanishing frequencies. The soft modes are analogous
to pure gauge degrees of freedom because one can add an arbitrary
number of these modes without changing the canonical energy $H$.
However, it is the energy $E$ which is the physically observable
quantity and, if $E$ is fixed, {then for given $H$} the number of the
soft modes cannot be arbitrary.

In what follows we interpret negative contributions of some of
the non-minimally coupled constituents to the entropy as the effect
of such ghost-like degrees of freedom. We adopt computations of
conformal field theories with central charges $c=1-6\xi$ in
order to obtain the distribution over the physical energies $E$
instead of the canonical distribution. Note that, like gauge
theories, the induced gravity models are unitary theories. The
difference between $E$ and $H$ appears only if a Cauchy surface is
divided by the horizon into "observable" and "non-observable"
regions. This difference is so important in the entropy problem
since the entropy of the black hole itself is connected with the
entanglement generated by the existence of the horizon. On the
entire Cauchy surface, $E$ and $H$ coincide.

\section{Counting the black hole entropy}
\setcounter{equation}0

The above arguments lead to the following conjecture:
i) Each induced gravity constituent with the momentum
$p$ and mass $m_k$
corresponds to a 2D conformal theory with
a central charge $c_k$ and a finite correlation length
$R(p)=|m_k({\bf p})|^{-1}$, $p=|{\bf p}|$;
charges of spinor constituents are
$c_d=2$, while charges of scalar fields are $c_s=1-6\xi_s$ and
depend on non-minimal couplings.
ii) Each constituent yields a contribution to the total
entropy equal to
\begin{equation}\label{2.24a}
s(c_k,R_k(p)/\epsilon)={c_k \over 6} \ln {R_k(p) \over \epsilon}~,
\end{equation}
where $\epsilon$ is some cutoff near the horizon, universal
for all fields. Equation (\ref{2.24a}) follows from (\ref{2.24}).

To proceed it is convenient
to represent the induced gravity constraints,
which eliminate divergences in
the Newton constant (see (\ref{1.2})) in the form
\begin{equation}\label{2.20}
C=\sum_s c_s+\sum_d c_d~,\hskip 1cm ~C=0~,
\end{equation}
\begin{equation}\label{2.21}
\sum_s c_sm_s^2+\sum_d c_dm_d^2=0~.
\end{equation}
Constant $C$ can be interpreted as a total central charge of constituents.
The charge is zero because at each momentum $p$
the 2D theory is free from ultraviolet
divergences.
Let us repeat that conditions (\ref{2.20}), (\ref{2.21})
are possible only when some
scalar central charges $c_s$ are negative.

Now it is easy to see that the entropy of all constituents in 2D
induced gravity at some momentum $p$ is
\begin{equation}\label{3.31}
s(p)=\sum_k s(c_k,R_k(p)/\epsilon)=
{1 \over 6} \sum_k c_k \ln R_k(p)={\pi \over G_2(p)}~,
\end{equation}
where $G_2(p)$ is the 2D induced Newton constant defined
in (\ref{1.18}). The dependence on cutoff $\epsilon$ disappears
because of  (\ref{2.20}). Note that we consider the
models with strictly positive couplings $G_2(p)$ and, therefore,
the  partial entropy $s(p)>0$, despite of contributions of some
negative central charges. One can also see that for a black hole
in 2D induced gravities (\ref{1.17}) the entropy (\ref{3.31})
coincides with the Bekenstein-Hawking entropy.
Furthermore, the entropy of 4D theory, which is
\begin{equation}\label{3.32}
S={a^2 \over 4\pi} \int_0^\infty s(p) dp^2={{\cal A} \over 4G}~,
\end{equation}
coincides with the Bekenstein-Hawking entropy (\ref{1.5})
of a four-dimensional
black hole with the horizon area ${\cal A}=a^2$.
The last equality in (\ref{3.32})
follows from  relation (\ref{1.21}) between 4D and 2D couplings.

Therefore, entropy of induced gravity constituents computed
in the near-horizon limit as the entropy of 2D conformal
fields does reproduce the Bekenstein-Hawking entropy.
This indicates that the approach by Carlip
\cite{Carlip:99}--\cite{Carlip:02} may have realization
in quantum gravity at the microscopic level.

The result (\ref{3.32}) can be also obtained in a slightly
different way.
Let us first find the total entropy
$S_k$ of each constituent by integrating over the transverse
momentum
\begin{equation}\label{3.4}
S_k={a^2 \over 4\pi} \int_0^\Lambda
s(c_k,R_k(p)/\epsilon)dp^2=
c_k{{\cal A} \over 48\pi}\left(m_k^2\ln {m_k^2 \over \Lambda}
+\Lambda(1-\ln\epsilon ^2 \Lambda )\right)
~,
\end{equation}
where $\Lambda$ is a upper cutoff on the momentum ${\bf p}^2$,
and then take
the sum of entropies of all constituents
\begin{equation}\label{3.5}
S=\sum_k S_k={{\cal A} \over 48\pi}\left(\sum_k c_km_k^2\ln m_k^2
-\ln \Lambda \sum_k c_k m_k^2
+ \Lambda(1-\ln\epsilon ^2 \Lambda )\sum_k c_k\right)=S^{BH}
~.
\end{equation}
Again the final result is finite and does not
depend on the cutoff $\Lambda$ because of
constraints (\ref{2.20}), (\ref{2.21}).

Our last comment concerns the energy associated with the entropy
of constituents. Fixing the energy is
important for approach \cite{Carlip:99}--\cite{Carlip:02}
because application of Cardy formula requires to know
the values of generators $L_0$ and $\bar{L}_0$ whose degeneracy
is studied.
The energy of each 2D constituent with
momentum $p$ can be found from the corresponding
free energy (\ref{2.26}).
For theory with a central charge $c_k$ one finds
\begin{equation}\label{3.8}
E(c_k,R_k(p)/\epsilon)=\left.\partial_\beta
(\beta
F(\beta,c_k,R_k(p)/\epsilon))\right|_{\beta=\beta_H}~.
\end{equation}
The result taken at the Hawking temperature (corresponding to
$\beta_H^{-1}=\kappa/(2\pi)$) is
\begin{equation}\label{3.9}
E(c_k,R_k(p)/\epsilon)=\frac 12 T_H s(c_k,R_k(p)/\epsilon)~~
\end{equation}
and the total 2D and 4D energies are
\begin{equation}\label{3.10a}
E(p)=\sum_k E(c_k,R_k(p)/\epsilon)=\frac 12 T_H s(p)
\end{equation}
\begin{equation}\label{3.10b}
E={a^2 \over 4\pi}
\int E(p)dp^2 =\frac 12 T_H S^{BH}~.
\end{equation}
Thus, the energy is proportional to the Bekenstein-Hawking entropy.

\section{Discussion}
\setcounter{equation}0

The aim of this paper was to investigate a role of the near-horizon
conformal symmetry in calculating the Bekenstein-Hawking
entropy $S^{BH}$ in induced gravity models. The arguments that
there is a universal way to  represent $S^{BH}$
in a statistical-mechanical form by using a near-horizon CFT
were given in \cite{Carlip:99}--\cite{Solo:98}. Being very
general these arguments do not take into account any
microscopic structure of the theory.
Therefore, the question
of whether the above approach can have a microscopic realization
remains an open problem. It is that issue which has motivated our work.
The microscopic degrees of freedom of induced gravity models
are free quantum fields obeying certain constraints.
So it is interesting to see how the ideas of
\cite{Carlip:99}--\cite{Solo:98} may or may not work
in this simple case.

We have demonstrated that induced gravity models in a region
near the horizon can be dimensionally reduced in such a way that
each 4D constituent field results in a tower of 2D quantum theories
defined
in a 2D plane orthogonal to the horizon. Close to the horizon
the 2D fields are effectively
massless and under coordinate
changes their stress-energy tensors transform as stress tensors
of 2D conformal theories, each with its own central charge.
This enables one to count the states of 2D induced gravities by using
standard CFT methods.  Summation of all partial entropies
$s(p)$ reproduces the correct value of  $S^{BH}$.
By its nature, this method is very
close to the idea of \cite{Carlip:99}--\cite{Solo:98}.  So
one can say that the latter does have a realization in
the induced gravity models.

However, several important features are  missing in the previous
works. Our first observation is that the total central charge in
near-horizon CFT in 2D induced gravities at each transverse momentum
$p$ vanishes. This property just reflects the absence of the
ultraviolet divergences in such theories. The second observation is
that a non-zero value of a partial  entropy $s(p)$ is a result of
breaking the conformal symmetry. The reason is that the constituents
are massive fields and CFT's are characterized by finite correlation
lengths. Moreover, because masses of fields do not coincide in
general, there is a variety of correlation lengths in the theory.

It is this feature which makes
our computation distinct from the approach of
\cite{Carlip:99}--\cite{Solo:98}.
A CFT in \cite{Carlip:99}--\cite{Solo:98}
is characterized by a single non-trivial central charge $c$ and a
single correlation length $l$. There is a freedom in choosing
$l$ or $c$. If $l$ is fixed then $c$ is fixed too.
The induced gravity suggests
that, perhaps, $l$ is not a free parameter but a scale
dynamically determined by quantum gravity effects at Planckian energies.

Our discussion leaves a number of open problems.
One of them is the physical meaning of negative
central charges which appear due to non-minimally coupled scalar
constituents.
Although we have argued that each partial
entropy $s(p)$ is positive,
there is a problem how to interpret the negative contributions
to this quantity.
One of the
possibilities mentioned in section 3.3
is that negative charge CFT's correspond to ghost-like degrees of
freedom.
Such degrees of freedom may appear on a part of the Cauchy hypersurface
when it is divided by the horizon but it
should not violate unitarity of the theory.
It is fair to say that further work on this issue
is needed.

In this paper we have reduced 4D induced gravity to
a set of two-dimensional theories which we considered as
fully physical ones.
The nice features of 2D theories is a possibility
to apply Zamolodchikov's $c$-theorem
\cite{Zamol},\cite{Cardy}. The theorem concerns an information loss
under renormalization transformations in the direction of larger
scales. It is interesting to search for possible relation between this information
loss and the entropy of a black hole in induced gravity.

To this aim let us return to results of sections 3.1, 3.2
and consider a quantum scalar field of mass $m$ on a flat
2D space-time. Suppose
we study  a correlator of this field (\ref{2.11}) in the
Rindler coordinates (\ref{2.3}). Put one point in the correlator
on the horizon $\rho=0$ and the other point
at some nonzero value $\rho=R$. We have two regimes: if $R \ll m^{-1}$
the correlator behaves as in massless theory, if  $R \gg m^{-1}$ it is
exponentially suppressed by the mass. One can interpret the evolution
along the Rindler radius $R$ as a renormalization group flow
\cite{Cardy} between the two fixed points, the ultraviolet point (UV),
$R=0$, and infrared point (IR), $R\rightarrow \infty$.
These asymptotic limits of the massive theory are described by UV
and IR scale-invariant theories which lie at the end points of
renormalization group trajectory, see, e.g.,  \cite{CFL}.
The UV central charge is $c_{UV}=1$ and it coincides with the central
charge of 2D massless scalar field theory, while the IR central
charge $c_{IR}$ vanishes.

Zamlodochikov's $c$-theorem enables one to construct
a $c$-function $C(R,m)$ which is a combination of local correlators
of components of the renormalized stress-energy tensor of the
field $\phi$,
such that $C(R,m)$ is non-increasing along renormalization
group trajectory, $\partial _R C(R,m)\leq 0$, see the details
in \cite{Cardy},\cite{CFL}. At the fixed points the $c$-function
coincides with the corresponding central charges. The mass of the
field can be interpreted as a coupling constant while the mass term
in the action near the fixed point $R=0$ can be
considered as a perturbation.
Zamolodchikov's theorem  indicates that there should
exist some kind of entropy which measures the loss of information
about the degrees of freedom of the theory whose wave length
is smaller than the cutoff $R$.

In section 4 we defined a partial entropy $s(p)$
of constituents in 2D induced
gravity as a sum of contributions from different fields.
Because some fields correspond to CFT's with negative
central charge
the application of the $c$-theorem to each individual field
is not justified because there is a problem with unitarity.
The 2D fields at a given $p$ should be considered only as a whole system.
Such system is well-defined (free
of ultraviolet divergences) and
can be unitary. Note that transformations changing the length scale
of the theory are equivalent to
redefinition of transverse energies of fields $m({\bf p})$.
This is easy to see by examining correlators of massive
fields (\ref{2.11}). However, if one considers trivial transformations
$m'({\bf p})=\alpha m({\bf p})$
the partial entropy $s(p)$ and 2D gravitational coupling $G_2(p)$
do not change, see (\ref{3.31}).

To have an analog of a renormalization flow in the
induced gravity models one can choose another option. Let us
consider the momentum $p$ as a renormalization parameter. This makes
sense for several reasons. First, decreasing $p$ means decreasing
transverse energies and is equivalent to going to larger length
scales of the theory. Second, by changing $p$ one moves from one 2D
induced gravity to another in a consistent way without violating the
induced gravity constraints. Third, as we discussed in the end of
section 2.2, the parameter $p$ becomes a real ultraviolet cutoff when
the induced gravity theory is considered on space-times (\ref{1.7a}),
(\ref{1.7b}) with 2D curvature ${\cal R}$ larger than $m^2_{Pl}+p^2$.

Identification of $p$ with the renormalization parameter has
interesting consequences. In the ultraviolet limit, when $p\gg m_{Pl}$,
4D and 2D running gravitational couplings, $G(p)$,
$G_2(p)$, are large. On the other hand, the partial entropy $s(p)$
and the Bekenstein-Hawking entropy vanish, in accord with the fact that
no information is lost in UV limit. In the infrared limit,
$p\rightarrow 0$,
the couplings are finite and the value of $G(p)$ approaches
the value of the Newton constant for 4D induced gravity
given by (\ref{1.4}).
The loss of the information in the IR limit is maximal
and it can be measured by the Bekenstein-Hawking entropy.
The coupling $G_2(p)$ can be related
to the $\beta$-function of the theory, see (\ref{1.22}).
If $G_2(p)>0$ the 4D running coupling $G(p)$
is monotonically increasing when one moves
for UV to IR points. Therefore, the parameter $G^{-1}(p)$
behaves as a $c$-function.

 To summarize, using the induced gravity as a toy model one can
obtain an interesting information concerning the mechanism of the
entropy generation in black holes. The entanglement of the
constituents in the presence of the horizon plays a fundamental role
in this mechanism. The universality of the Bekenstein-Hawking
entropy, that is its independence of the detailed structure of the
background microscopic theory of constituents which induce the
gravitational field, is a natural property of the induced gravity
models. In this paper we demonstrated that the induced gravity
models  allow one to exploit the results of the CFT for the
explanation of  black hole entropy. This can be used to check
different ideas proposed in this area by the string theory.  We are
fully aware of the remaining difficulties which indicate that even in
such simplified models the statistical-mechanical calculation of the
Bekenstein-Hawking entropy is a very non-trivial task.

\bigskip
\vspace{12pt} {\bf Acknowledgements}:\ \

This work was partially supported by the Natural Sciences and
Engineering Research Council of Canada. The authors would like to
thank the Killam trust for its support.

\appendix

\section{Examples of induced gravity models with steady RG flow}
\setcounter{equation}0

Explicit examples of induced gravity models with two-dimensional
constant $G_2(p)$ positive for all $p>0$ is easy
to construct. Not to have problems with induced cosmological constant
consider a version of induced gravity where all
constituents are combined in supersymmetric multiplets.
Each multiplet consists of one Dirac spinor and 4 real scalar fields.
The masses of of all fields entering the same multiplet coincide.
Also non-minimal coupling constants of scalar components
insides the
multiplets are the same. Suppose we have $N$ multiplets (thus, $N=N_d$)
and let $m_j$, $\xi_j$ be mass and non-minimal coupling of the
$j$th multiplet. Constraints (\ref{1.1}) in such a theory
are satisfied and the induced cosmological constant vanishes.
Relations (\ref{1.2}), (\ref{1.4}) can be written in the
form
\begin{equation}\label{b.1}
\sum_{j=1}^{N}x_j=0~,\hskip 1cm
\sum_{j=1}^{N}x_jm_j^2=0~,\hskip 1cm
\sum_{j=1}^{N}x_j m_j^2\ln m_j^2={\pi \over G}~,
\end{equation}
where $x_j=1/2-2\xi_j$.
One can consider (\ref{b.1}) as a system of equations
which defines parameters $x_j$ at given masses $m_j$ and a given
value $G$. The system is linear and allows solutions for
$N\geq 3$.

The solution is easy to find for $N=3$ and to
get explicit expressions for 2D gravitational
coupling $G_2(p)$ and the running 4D
coupling $G(p)$, see (\ref{1.18}), (\ref{1.21}).
After some algebra the results can be represented
in  a simple form
\begin{equation}\label{b.4}
{G \over G(p)}={D(p) \over D(0)}~,\hskip 1cm
{G_2(0) \over G_2(p)}={D'(p) \over D'(0)}~,
\end{equation}
\begin{equation}\label{b.5}
D(p)=f(\mu_1(p),\mu_2(p),\mu_3(p))~,\hskip 1cm
\end{equation}
\begin{equation}\label{b.5a}
f(a,b,c)=(c-b)a\ln a+
(a-c)b\ln b +
(b-a)c\ln c
~,
\end{equation}
\begin{equation}\label{b.6}
D'(p)={\partial \over \partial p^2} D(p)=g(\mu_1(p),\mu_2(p),\mu_3(p))~,
\end{equation}
\begin{equation}\label{b.6a}
g(a,b,c)=(c-b)\ln a+
(a-c)\ln b +
(b-a)\ln c
~,
\end{equation}
where $\mu_k(p)=m_k^2+p^2$ and $m_1$, $m_2$, $m_3$
are the masses of the three multiplets. The 2D coupling is
defined as
\begin{equation}\label{b.7}
{1 \over G_2(0)}=-{\partial \over \partial p^2} {1 \over
G(p)}_{p=0}
\end{equation}
It can be shown that if $G>0$ than both functions
$G(p)$, $G_2(p)$ are positive and monotonically
increasing as parameter $p$ increases.

\section{Black hole entropy in 2+1 dimensions}
\setcounter{equation}0

In this Appendix we demonstrate that our derivation of the black hole
entropy works space-times where the number of dimensions
is different from four. As an example, we consider
a theory in a three dimensional space-time. This case is instructive because
quantum theories in odd and even dimensions have different properties.
It is also interesting because the entropy of some three-dimensional
black holes, such as the BTZ  (Banados-Teitelboim-Zanelli) black
hole \cite{BTZ}, allows statistical-mechanical representation
by methods of conformal theory \cite{Strominger}.

Induced gravity models in three dimensions were discussed
in \cite{FF:99ch}. For the model consisting of spinor and
non-minimally coupled scalar  constituents the constraints
which ensure finiteness of the induced Newton constant
can be written in the form (\ref{2.20}). Central charges
of scalar fields are $c_s=1-6\xi_s$, while charges
of spinor fields are $c_d=1$ in accord with the fact that spinors
have 2 components in $D=3$. Given these constraints the induced
Newton constant is defined by the relation \cite{FF:99ch}
\begin{equation}\label{a.1}
{1 \over G}=-\frac 13 \sum_k c_k m_k~,
\end{equation}
where the parameters $\xi_s$ are chosen so that $G>0$.
The horizon of a black hole in three dimensions is a circle
of some length $a$ \cite{BTZ} and the corresponding Bekenstein-Hawking
entropy is
\begin{equation}\label{a.2}
S^{BH}={a \over 4G}~.
\end{equation}
We now use the same line of arguments as in four dimensions
and consider entropy of constituents in the near-horizon limit.
In the limit when $a$ is large
the total entropy of 2D constituents at the momentum $p$
coincides with (\ref{3.31})
\begin{equation}\label{a.3}
s(p)=\sum_k s(c_k,R_k(p)/\epsilon)=
{1 \over 6} \sum_k c_k\ln R_k(p)~.
\end{equation}
The entropy of the 3D theory is
\begin{equation}\label{a.4}
S={a \over \pi}\int_{0}^{\infty}
s(p) dp~.
\end{equation}
By performing the
integration  in (\ref{a.4}) one finds
the expression
\begin{equation}\label{a.6}
S=-{a \over 12} \sum_k c_k m_k~,
\end{equation}
which is finite because of the induced gravity constraints
and coincides with the Bekenstein-Hawking entropy (\ref{a.2}).

\end{document}